\documentclass[letterpaper]{article} 
\usepackage{aaai2026}
\usepackage{times}  
\usepackage{helvet}  
\usepackage{courier}  
\usepackage[hyphens]{url}  
\usepackage{graphicx} 
\usepackage{natbib}  
\usepackage{caption} 
\usepackage{algorithm}
\usepackage{algorithmic}

\usepackage{latexsym}
\usepackage{inconsolata}
\usepackage{graphicx}
\usepackage{enumitem}
\usepackage{makecell}
\usepackage{multirow}
\usepackage{subfigure}
\usepackage{framed}
\usepackage{mdframed}
\usepackage{booktabs}
\usepackage{amsmath}
\usepackage{soul}
\usepackage{pdflscape}
\usepackage[HTML]{xcolor}

\usepackage{newfloat}
\usepackage{listings}
\DeclareCaptionStyle{ruled}{labelfont=normalfont,labelsep=colon,strut=off} 
\floatstyle{ruled}
\newfloat{listing}{tb}{lst}{}
\floatname{listing}{Listing}
\title{ARMove: Learning to Predict Human Mobility through Agentic Reasoning}
\author{
    Chuyue Wang\textsuperscript{\rm 1},
    Jie Feng\textsuperscript{\rm 2}$^*$,
    Yuxi Wu\textsuperscript{\rm 3},
    Shenglin Yi\textsuperscript{\rm 4}, 
    Hang Zhang\textsuperscript{\rm 4}
}
\affiliations{

    \textsuperscript{\rm 1}Tencent\\
    \textsuperscript{\rm 2}Zhongguancun Academy\\
    \textsuperscript{\rm 3}Alibaba Group\\
    \textsuperscript{\rm 4}Meituan\\
    fengj12ee@hotmail.com\\

}

\usepackage{bibentry}

\begin{document}

\maketitle

\begin{abstract}
Human mobility prediction is a critical task but remains challenging due to its complexity and variability across populations and regions. Recently, large language models (LLMs) have made progress in zero-shot prediction, but existing methods suffer from limited interpretability (due to black-box reasoning), lack of iterative learning from new data, and poor transferability. In this paper, we introduce \textbf{ARMove}, a fully transferable framework for predicting human mobility through agentic reasoning. To address these limitations, ARMove employs standardized feature management with iterative optimization and user-specific customization: four major feature pools for foundational knowledge, user profiles for segmentation, and an automated generation mechanism integrating LLM knowledge. Robust generalization is achieved via agentic decision-making that adjusts feature weights to maximize accuracy while providing interpretable decision paths. Finally, large-small model synergy distills strategies from large LLMs (e.g., 72B) to smaller ones (e.g., 7B), reducing costs and enhancing performance ceilings. Extensive experiments on four global datasets show ARMove outperforms state-of-the-art baselines on 6 out of 12 metrics (gains of 0.78\% to 10.47\%), with transferability tests confirming robustness across regions, users, and scales. The other 4 items also achieved suboptimal results. Transferability tests confirm its 19 robustness across regions, user groups, and model scales, while interpretability 20 analysis highlights its transparency in decision-making. Our codes are available at: \url{https://anonymous.4open.science/r/ARMove-F847}.
\end{abstract}

\section{Introduction}
Mobility prediction task is widely used across multiple disciplines and fields. For instance, it can be used to predict travel behavior in transportation to support the traffic condition forecasting. It can also play an important role in infectious disease modeling by providing data-driven support for public health policy-making.

Research on mobility prediction has a long history. In recent years, the field has primarily focused on deep learning methods, such as DeepMove~\cite{feng2018deepmove} and GETNext~\cite{Yang_2022}. These methods optimize model weights in a data-driven manner, enabling the models to discover and memorize mobility patterns from trajectory for predicting future movements. While such methods have achieved significantly better performance than earlier physical modeling approaches due to their ability to fit complex patterns in the trajectory data, they are subject to notable drawbacks, including a high risk of overfitting, limited generalization capacity, and a lack of interpretability.

Recently, large language models (LLMs) have demonstrated remarkable sequence modeling and reasoning generalization capabilities, leading to widespread application in various domains. Some researchers~\cite{wang2023would, feng2025agentmovelargelanguagemodel} have attempted to apply LLMs to mobility behavior prediction and have achieved promising results, particularly in terms of transferability and generalization. For example, AgentMove~\cite{feng2025agentmovelargelanguagemodel} can achieve zero-shot prediction without specific training and delivers performance comparable to that of well-trained deep learning models. However, AgentMove lacks the ability for iterative learning, meaning it cannot directly improve through incremental ingestion of new trajectory data. Furthermore, its prediction process depends on the black-box mechanisms of LLMs, resulting in limited interpretability.

This paper introduces ARMove, a novel agent-based framework for interpretable and sustainable mobility prediction via reasoning. ARMove addresses existing limitations by integrating three core modules. First, the Feature Module systematically summarizes mobility regularity, creating a standard feature pool with an augmentation mechanism. Coupled with an iterative process, this module efficiently mines mobility features from limited trajectory data through feedback. Second, the User Classification Module clusters users based on their mobility patterns, enabling more precise feature matching and iterative combination tailored to specific user groups. Finally, the Model Knowledge Transfer mechanism leverages feature weights to transfer learned feature combinations and user grouping strategies from a Large Language Model (LLM) to smaller LLMs. This collaboration between models effectively reduces the prediction burden while improving the performance ceiling of smaller models. Quantitatively and qualitatively validated through experiments, ARMove's linked modules facilitate learning from new data, enhance generalization via user clustering and weight-based transfer, and provide explicit feature reasoning paths for strong interpretability.

In summary, our contributions are as follows:
\begin{itemize}[leftmargin=*]
    \item We propose ARMove, the first agent-based framework applying agentic reasoning to human mobility prediction for enhanced efficiency and interpretability.
    \item We design three interdependent core modules: an Iterative Feature Module, a User Classification Module, and a Model Knowledge Transfer Mechanism based on feature weights.
    \item We pioneer a Large-to-Small LLM collaboration strategy that transfers learned knowledge from a large model to guide smaller prediction models, significantly reducing computational overhead while boosting the performance of the smaller models.
    \item ARMove achieves strong interpretability via explicit reasoning paths and superior adaptability to new data, validated by extensive experiments.
\end{itemize}
\section{Methods} \label{sec:method}
\subsection{Overview}
\begin{figure*}
  \centering
  \includegraphics[width=\textwidth]{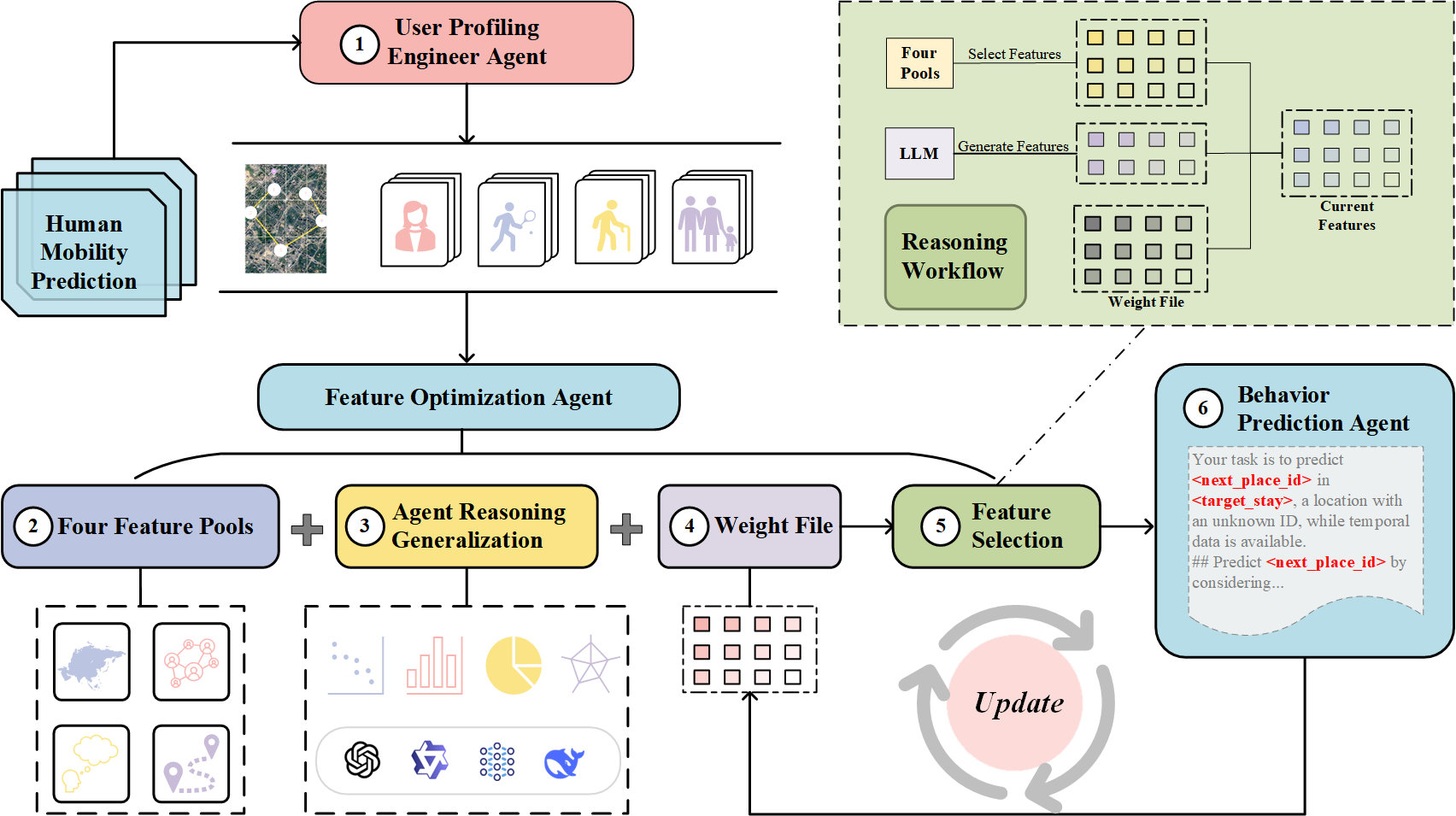}
  \caption{The framework of ARMove.}
  \label{fig:main-pic}
\end{figure*}
As shown in Figure~\ref{fig:main-pic}, ARMove consists of three main agent collaboration modules: the Feature Optimization Agent, the User Profiling Engineer Agent, and the Behavior Prediction Agent. The Feature Optimization Agent is primarily responsible for feature standardization, feature selection, and feature generation. The User Profiling Engineer module mainly controls the multi-level generation of user profiles and iterative collaboration with the Feature Optimization Agent. The Behavior Prediction Agent is mainly responsible for generating the final human mobility predictions. In addition, compared to existing large model methods, we propose model transfer along three dimensions: model, user, and region.

\subsection{Feature Optimization Agent}
Traditional large models have relied on manual feature engineering for human mobility. This approach, while sometimes locally accurate, lacks generalizability and fails to leverage the model's internal knowledge.
Moreover, these hand-crafted features are often incomplete or redundant.
We therefore propose a novel Feature Optimization Agent. It dynamically loads and generates features, simultaneously eliminating redundancy and leveraging the model's internal knowledge. The objective is to unearth a more diverse feature set to boost predictive accuracy potential.

\textbf{Four Standard Feature Pools. } 
To enable multi-dimensional mobility prediction, we established a foundational feature pool, manually curated and refined from AgentMove. It covers four main categories (and 16 subcategories):

\begin{itemize}[leftmargin=*]
\item \textbf{Trajectory:} Includes trajectory time, context stay counts, target stay duration, historical visit frequency, and major venue IDs.
    
\item \textbf{Spatial:} Extracted via the Nominatim API, this covers administrative area information, subdistrict counts, and POI collections.
    
\item \textbf{Memory:} Long- and short-term memory, top activity times, recent visits, and user profile keywords to enable personalization.
    
\item \textbf{Social:} Encompasses each user's social network and neighbor networks constructed from raw social information and hop-level features.
\end{itemize}

This comprehensive pool provides a structured feature representation for each user, offering a solid data foundation for subsequent selection and prediction tasks.

\textbf{New Feature Generation. }While the standard feature pool offers strong generalizability, it falls short of enabling personalized modeling for diverse user groups, as a single feature set often fails to capture crucial variations in user behaviors and scenarios.
To address this, we propose a novel feature generation mechanism. This mechanism allows the agent to generate and compute diverse, scenario-specific features directly from user trajectory information. By processing the standard features, the generation module automatically reasons about potentially missing feature dimensions, mitigating the inherent trade-off between redundancy and omission found in manually designed features.

\textbf{Feature Selection.}
Following feature generation, we employ multi-round iterative training to select the optimal features for users in the current city. A key component is the introduction of a feature weight maintenance phase, which dynamically adjusts feature importance, prioritizing those better suited to the current urban environment to counteract the inherent randomness of selection.
Instead of directly maximizing the non-differentiable $\mathrm{Acc@k}$ metrics, the objective of the iterative optimization is to maximize a composite objective function $J$, which is a weighted combination of $\mathrm{Acc@1}$ and $\mathrm{Acc@5}$. This function serves as the proxy loss guiding the weight adjustment process. The final selection result is a comprehensive balance among agent-based selection, newly generated features ($\mathcal{F}_{\text{new}}$), and high-weight features ($\mathcal{F}_{\text{high}}$), built upon the standard feature set ($\mathcal{F}_{\text{std}}$).

\begin{equation*}
\label{eq:feature_select}
J(\mathbf{w}, \mathcal{F}) = \lambda \cdot \mathrm{\textit{Acc@1}}(\mathcal{F} \mid \mathbf{w}) + (1-\lambda) \cdot \mathrm{\textit{Acc@5}}(\mathcal{F} \mid \mathbf{w})
\end{equation*}

\subsection{User Profiling Engineer Agent}
After feature selection, we observe significant behavioral disparities among users in the same city. Modeling features for each individual user wastes resources, risks overfitting, and causes poor generalization. Thus, we propose the innovative User Profiling Engineer Agent. This agent infers user profile labels from trajectory features, enabling users to be grouped by profile for efficient human mobility prediction.

\textbf{User Grouping Strategy.}
The agent first uses the four standard feature pools to assign a baseline persona label (e.g., office worker) to each user, ensuring interpretability. However, initial categories often become too broad or too fragmented.
To address this, we introduce a reasoning approach to explore secondary user interests and dynamically merge similar groups. For example, users with the same primary persona may differ in secondary interests like fitness or nightlife. Identifying these patterns and consolidating redundant groups creates a more accurate and efficient user representation. This multi-level, dynamic profiling adapts the framework to diverse user groups and regions, supporting personalized prediction.

\textbf{Interaction with Feature Selection.}
User grouping shifts the core challenge: how to collaborate with the Feature Optimization Agent. Feature selection and new feature generation are now conducted at the user group level, generating group-specific features.

We maintain a global feature weight to prevent overfitting to any single group. The group category itself is also treated as a new feature for iterative reasoning. 
The optimization objective is performed for each group $g$ by maximizing a composite objective function $J_g$. This function serves as the proxy loss guiding the group's specific feature selection and weight adjustment process. 
The composite objective function for user group $g$ is defined as:
\begin{equation*}
\label{eq:group_feature_select}
J_g(\mathbf{w}_g, \mathcal{F}_g) = \lambda \cdot \mathrm{\textit{Acc@1}}_g(\mathcal{F}_g \mid \mathbf{w}_g) + (1-\lambda) \cdot \mathrm{\textit{Acc@5}}_g(\mathcal{F}_g \mid \mathbf{w}_g)
\end{equation*}
Here, $\mathcal{F}_g$ is the total feature set for group $g$, which includes standard, group-specific new, and high-weight features, as well as the group label. $\mathbf{w}_g$ is the dynamically adjusted feature weight vector specific to group $g$. $\lambda \in [0, 1]$ is the hyperparameter balancing $\mathrm{Acc@1}$ and $\mathrm{Acc@5}$.

\subsection{Model Transfer}
Since many previously proposed LLM-based frameworks for human mobility prediction do not involve a training process, they also lack the capability for knowledge transfer. In our proposed ARMove framework, we innovatively introduce three types of transfer methods: collaborative transfer between large and small models, transfer between different users within same city, and transfer between different cities.

\textbf{Large-Small Model Collaboration Strategy.}
To mitigate the significant computational and financial costs of multi-agent collaboration, we propose a novel large model guiding small model strategy. We design a user profile reuse mechanism and a city weight reuse strategy that directly transfer the knowledge, user profiles, and feature weights learned by large models (GPT-4o-mini) to smaller models (8B models). This mechanism allows smaller models to efficiently inherit the extensive, multi-city knowledge and reasoning capabilities of the large models, substantially boosting their predictive performance while significantly reducing computational resource consumption and cost. Consequently, the small models' performance closely approaches that of the large models, greatly enhancing the system's efficiency and scalability for large-scale mobility prediction.

\textbf{Same-City,Different-Users.}
In the same-city, different-user transfer setting, we sample user data from the original city and randomly replace a subset of users. This allows us to evaluate the effectiveness and generalizability of training data within the same city. For the newly replaced users, we perform user grouping and feature optimization as usual. However, in this scenario, the feature weight updates during optimization are initialized or replaced with the results obtained from the training process of the original users in the same city. This approach ensures that the feature optimization process benefits from previously learned knowledge, while still adapting to the characteristics of the new user group.

\textbf{Transfer Between Different Cities.}
Inter-city transfer is evaluated using two primary methods: multi-city fusion transfer and direct transfer between individual cities.
Multi-city fusion transfer involves constructing a comprehensive global human trajectory dataset by randomly sampling data across multiple cities. This fused dataset integrates diverse geographic, cultural, and behavioral characteristics, increasing complexity. While this complexity poses challenges for user grouping and model generalization, it enables the model to learn more universal human mobility patterns.
Direct inter-city transfer applies the trained model parameters or feature weights from one source city directly to a target city's dataset. This approach assesses whether knowledge transfer improves performance in target cities with similar characteristics and evaluates the model's adaptability to differences in geography and user behavior.
If a diverse (fused) or single-city dataset can be successfully transferred to an untrained target city, achieving performance comparable to a locally trained model or maintaining stable results, it demonstrates the strong effectiveness and practical value of the transfer strategy. This outcome not only enhances model scalability but also provides methodological support for large-scale, cross-regional mobility prediction in real-world applications.

\section{Evaluation} \label{sec:eval}
\subsection{Settings}
\subsubsection{Datasets}
\noindent\textbf{Location Representation.} 
In the Foursquare datasets, each location is represented by its unique 
POI (Point of Interest) ID from the Foursquare venue database. 
For the ISP dataset from Shanghai, raw GPS coordinates are mapped 
to a uniform grid of 1km$\times$1km cells, resulting in discrete location IDs.

\noindent\textbf{Dataset Preprocessing and Evaluation Split.} 
Following the preprocessing in AgentMove~\cite{feng2025agentmovelargelanguagemodel}, 
we divide trajectories into sessions using a 72-hour window (Foursquare) or merge points by day (ISP). Sessions with fewer than 4 stays and users with fewer than 5 sessions are filtered. For fair comparison with deep learning baselines, we adopt a global temporal split: Foursquare data is split into train/validation/test sets in a 7:1:2 ratio; ISP data in 4:1:5. However, as ARMove operates in a zero-shot setting, no training is performed---all historical trajectories are used to construct prompts 
for feature extraction, user clustering, and knowledge transfer. 

\noindent\textbf{Training Paradigm.} 
ARMove is a zero-shot framework: no end-to-end training 
or parameter optimization occurs. Instead, historical trajectories 
(from the full dataset) are aggregated into prompts for the Feature 
Optimization Agent (standard pools, new generation, selection) and 
User Profiling Agent (clustering). Iterative refinement uses feedback 
from a small validation subset of prompts. The Behavior Prediction 
Agent generates zero-shot predictions on test trajectories using the 
finalized features and LLM reasoning.

\subsubsection{Baselines}
We compare our proposed model with the following baselines: \textbf{FPMC}~\cite{rendle2010factorizing},\textbf{RNN}~\cite{feng2018deepmove},\textbf{DeepMove}~\cite{feng2018deepmove}, \textbf{LSTPM}~\cite{sun2020go}, \textbf{GETNext}~\cite{Yang_2022},\textbf{STHGCN}~\cite{yan2023spatio}) and three LLM-based methods(\textbf{LLM-Mob}~\cite{wang2023would},\textbf{LLM-ZS}~\cite{beneduce2024large}, \textbf{LLM-Move}~\cite{feng2024nextzeroshotgeneralizationllms}, \textbf{AgentMove}~\cite{feng2025agentmovelargelanguagemodel}. 

\subsection{Evaluation Metrics}
In the experiments, we adopt the widely used evaluation metrics Accuracy@1, Accuracy@5, and NDCG@5 as the main evaluation criteria~\cite{sun2020go,luca2021surveydeeplearninghuman}.
\begin{itemize}[leftmargin=*]
    \item \textbf{Acc@k}: Accuracy at top-k, i.e., whether ground truth is in top-k predictions. 
    Measures exact match performance.
    \item \textbf{NDCG@5}: Normalized Discounted Cumulative Gain at 5, 
    which assigns higher scores to relevant items ranked higher. 
    Particularly important for mobility tasks where near-misses 
    (e.g., nearby POI) are partially correct.
\end{itemize}
All metrics are computed on the 200 sampled test trajectories per city.

\subsubsection{Implementation}
The implementations of FPMC, RNN, DeepMove, LSTPM, GETNext, STHGCN, and AgentMove are all based on the official codes provided by their respective authors. During model training and inference, we follow the default parameter settings specified in the libraries and official codes. For LLMs, we use the OpenAI API to access GPT-4o-mini, and utilize DeepInfra and OpenRouter to access other open-source LLMs. Detailed parameter settings for all baselines can be found in the appendix.

\subsection{Main Results}
We compare ARMove with 10 baseline methods on datasets from four cities, as shown in Table~\ref{table:main result}, using GPT-4o-mini as the base LLM for all LLM-based methods.
Among deep learning methods, GETNext and STHGCN perform best on several metrics, demonstrating strong modeling capabilities. Among LLM-based approaches, AgentMove is the most competitive baseline, surpassing deep learning methods on half of the metrics. 
Compared with all baselines, ARMove achieves the best results on 6 out of 12 metrics and the second-best on 3 metrics, ranking as the overall top performer. Among all deep learning methods, although ARMove underperforms GETNext on two metrics, it achieves improvements ranging from 1.66\% to 43.42\% over the best deep learning method on other metrics across the remaining datasets. Notably, ARMove outperforms AgentMove in 10 out of 12 metrics, with gains ranging from 1.35\% to 18.06\%. These results demonstrate that ARMove effectively leverages the multi-agent collaborative reasoning framework to significantly enhance human mobility prediction.

\begin{table*}
\centering
\caption{Main results of baselines and ARMove. All LLM-based methods use GPT4omini as the base LLM. Deep learning models are trained on each city's training set, while LLM-based models are evaluated on the test set in a zero-shot setting.}
\label{table:main result}
\resizebox{\textwidth}{!}{%
\renewcommand{\arraystretch}{1.2} 
\begin{tabular}{lcccccccccccc}
\Xhline{0.7pt} 
\hline
\textbf{Models} & \multicolumn{3}{c}{\textbf{Shanghai(ISP})} & \multicolumn{3}{c}{\textbf{Moscow}} & \multicolumn{3}{c}{\textbf{Tokyo}} & \multicolumn{3}{c}{\textbf{Saopaulo}} \\
 & \textbf{Acc@1} & \textbf{Acc@5} & \textbf{NDCG@5} & \textbf{Acc@1} & \textbf{Acc@5} & \textbf{NDCG@5} & \textbf{Acc@1} & \textbf{Acc@5} & \textbf{NDCG@5} & \textbf{Acc@1} & \textbf{Acc@5} & \textbf{NDCG@5} \\ \hline
\Xhline{0.3pt} 
\textbf{FPMC} & 0.130 & 0.355 & 0.249 & 0.020 & 0.065 & 0.043 & 0.060 & 0.165 & 0.121 & 0.045 & 0.085 & 0.066 \\
\textbf{RNN} & 0.065 & 0.175 & 0.123 & 0.090 & 0.185 & 0.140 & 0.105 & 0.240 & 0.176 & 0.095 & 0.230 & 0.169 \\
\textbf{DeepMove} & 0.175 & 0.320 & 0.251 & 0.165 & 0.335 & 0.258 & 0.175 & 0.320 & 0.251 & 0.150 & 0.310 & 0.236 \\
\textbf{LSTPM} & 0.095 & 0.17 & 0.135 & 0.140 & 0.255 & 0.196 & 0.145 & 0.280 & 0.218 & 0.190 & 0.365 & 0.281 \\
\textbf{GETNext} & 0.115 & 0.260 & 0.178 & 0.175 & \underline{0.380} & \underline{0.269} & \textbf{0.205} & 0.450 & 0.317 & 0.165 & \textbf{0.398} & 0.299 \\
\textbf{STHGCN} & 0.125 & 0.277 & 0.195 & \underline{0.180} & 0.372 & 0.265 & \underline{0.198} & 0.430 & 0.300 & 0.175 & \textbf{0.398} & 0.299 \\ \hline
\textbf{LLM-Mob} & 0.120 & 0.360 & 0.239 & 0.085 & 0.250 & 0.168 & 0.050 & 0.215 & 0.133 & 0.115 & 0.275 & 0.200 \\
\textbf{LLM-ZS} & 0.155 & 0.425 & 0.291 & 0.125 & 0.350 & 0.240 & 0.175 & 0.450 & 0.318 & 0.150 & 0.355 & 0.257 \\
\textbf{LLM-Move} & 0.135 & 0.425 & 0.3141 & 0.160 & 0.275 & 0.235 & 0.175 & 0.395 & 0.310 & \textbf{0.230} & 0.375 & \textbf{0.321} \\
\textbf{AgentMove} & \underline{0.210} & \underline{0.445} & \underline{0.338} & 0.155 & 0.370 & 0.263 & 0.160 & \textbf{0.475} & \textbf{0.323} & \underline{0.215} & 0.370 & 0.296 \\ \hline
\textbf{ARMove} & \textbf{0.232} & \textbf{0.477} & \textbf{0.360} & \textbf{0.183} & \textbf{0.383} & \textbf{0.293} & 0.170 & \underline{0.455} & \underline{0.320} & 0.200 & \underline{0.390} & \underline{0.300} \\
\textbf{vs AgentMove} & 10.47\% & 7.1\% & 6.50\% & 18.06\% & 3.51\% & 11.40\% & 6.25\% & -4.39\% & -0.93\% & -7.5\% & 5.40\% & 1.35\% \\
\textbf{vs Deep Learning} & 32.57\% & 34.36\% & 43.42\% & 1.66\% & 0.78\% & 8.92\% & -17.07\% & 1.11\% & 6.66\% & 5.26\% & -2.05\% & 3.44\% \\
\textbf{vs Best Baselines} & 10.47\% & 7.1\% & 6.50\% & 1.66\% & 0.78\% & 8.92\% & -17.07\% & -4.39\% & -0.93\% & -15.00\% & -2.05\% & -7.00\% \\ \hline
\Xhline{0.7pt} 
\end{tabular}%
}
\end{table*}

\subsection{Ablation Studies}
In this section, we conduct a detailed analysis of the different modules in the ARMove model to further verify the effectiveness of each component. We perform ablation experiments by removing each module individually and observing its impact on the overall performance. The relevant experimental results are shown in Table~\ref{table:Ablation pic}. 

\textbf{LLM New Feature Module.} LLM new feature also enhances model performance to some extent. For example, in the Shanghai dataset, after removing this module, Acc@1 increases to 0.232, but Acc@5 and NDCG@5 decrease to 0.46 and 0.36, respectively, indicating that this module plays a certain role in improving overall ranking quality. The removal of the Iteration mechanism has a relatively small impact on performance, but in the Tokyo and SaoPaulo datasets, Acc@1 increases to 0.184 and 0.175, suggesting that there may be room for optimization in some scenarios.

\textbf{Iteration.} Removing the iteration module led to a decrease in the Acc@1 metric in three out of four cities. This confirms the critical role of the iterative mechanism in human mobility prediction. By enabling the model to adjust feature weights and prediction strategies across multiple optimization rounds, the iteration module significantly enhances the accuracy of the initial prediction (Acc@1). These results underscore the necessity of iterative optimization for modeling complex human mobility behaviors.

\textbf{User Classify Module.} It is particularly effective in improving the Acc@1 metric. For example, in the Shanghai, Tokyo, and Moscow datasets, when the User Classify module is removed, Acc@1 drops from 0.226, 0.21, and 0.180 to 0.225, 0.21, and 0.180, respectively, indicating that this module makes a significant contribution to top-1 prediction. However, the improvements in Acc@5 and NDCG@5 brought by the User Classify module are relatively limited, and in some datasets (such as Moscow and SaoPaulo), there is even a slight decline, suggesting its limited benefit for long-tail recommendations.

\begin{figure*}[h!]
  \centering
  \includegraphics[width=\textwidth]{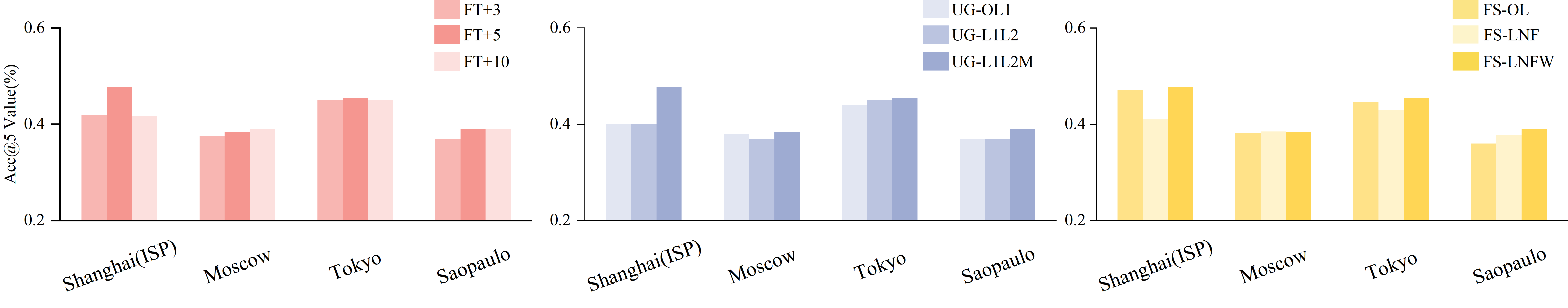}
  \caption{Performance acc@5 with custom legends.`FT+3’, `FT+5’, and `FT+10’ represent 3, 5, and 10 iterations, respectively;  `UG-OL1’, `UG-L1L2’, and `UG-L1L2M’ denote initial user profiling, user interest mining,and fine-grained category merging, respectively;  `FS-OL’, `FS-LNF’, and `FS-LNFW’ indicate feature selection by the large model only, feature selection by the large model combined with new feature generation, and balanced weighting.}
  \label{fig:3-1-XIAORONG_pic}
\end{figure*}

\textbf{Feature Selection.} It is worth noting that the Feature Selection (FS, AgentMove) module has a significant impact on overall performance. In all datasets, removing this module leads to a noticeable drop in Acc@1, Acc@5 and NDCG@5. For example, in the Shanghai dataset, Acc@1 drops from 0.226 to 0.210 and NDCG@5 drops from 0.36 to 0.3387, fully demonstrating the fundamental role of the feature selection module in the model. 
In addition to the ablation studies on individual modules, we also conducted in-depth ablation experiments on the internal mechanisms of certain modules, as illustrated in Figure~\ref{fig:3-1-XIAORONG_pic}. Specifically, we examined the effects of different iteration counts, user grouping strategies, and feature selection methods.

\begin{table*}
\centering
\caption{Ablation studies of ARMove. `- LLM New Feature' denotes removing the large language model (LLM) generated new feature module, `- Iteration' denotes removing the iterative module, `- User Classify' denotes removing the user classification module, and `- FS' denotes removing the feature selection module.}
\label{table:Ablation pic}
\resizebox{\textwidth}{!}{%
\renewcommand{\arraystretch}{1.2}
\begin{tabular}{lccccccccccccc}
\Xhline{0.7pt} 
\hline
\textbf{Modules} & \multicolumn{3}{c}{\textbf{Shanghai(ISP})} & \multicolumn{3}{c}{\textbf{Moscow}} & \multicolumn{3}{c}{\textbf{Tokyo}} & \multicolumn{3}{c}{\textbf{Saopaulo}} \\
 & \textbf{Acc@1} & \textbf{Acc@5} & \textbf{NDCG@5} & \textbf{Acc@1} & \textbf{Acc@5} & \textbf{NDCG@5} & \textbf{Acc@1} & \textbf{Acc@5} & \textbf{NDCG@5} & \textbf{Acc@1} & \textbf{Acc@5} & \textbf{NDCG@5} \\ \hline
\Xhline{0.3pt} 
\textbf{ARMove} & {\ul 0.226} & \textbf{0.477} & \textbf{0.360} & {\ul 0.176} & \textbf{0.383} & \textbf{0.293} & 0.170 & {\ul 0.455} & {\ul 0.320} & 0.200 & {\ul \textbf{0.390}} & {\ul \textbf{0.300}} \\
\textbf{- LLM New Feature} & \textbf{0.232} & 0.460 & 0.360 & \textbf{0.183} & 0.372 & 0.280 & 0.175 & 0.410 & 0.300 & 0.200 & 0.389 & 0.300 \\
\textbf{- Iteration} & 0.226 & 0.472 & 0.360 & 0.178 & 0.382 & 0.280 & 0.184 & 0.446 & 0.320 & 0.175 & 0.360 & 0.270 \\
\textbf{- User Classify} & 0.225 & 0.450 & 0.350 & 0.180 & 0.355 & 0.270 & {\ul \textbf{0.210}} & {\ul 0.460} & {\ul \textbf{0.340}} & 0.200 & 0.360 & 0.290 \\
\textbf{- FS} & 0.210 & 0.445 & 0.3387 & 0.155 & {\ul 0.370} & {\ul 0.263} & 0.160 & \textbf{0.475} & 0.323 & {\ul \textbf{0.215}} & 0.370 & 0.296 \\ \hline
\Xhline{0.7pt} 
\end{tabular}%
}
\end{table*}

\begin{figure}[htbp]
  \centering
    \centering
    \includegraphics[width=0.45\textwidth]{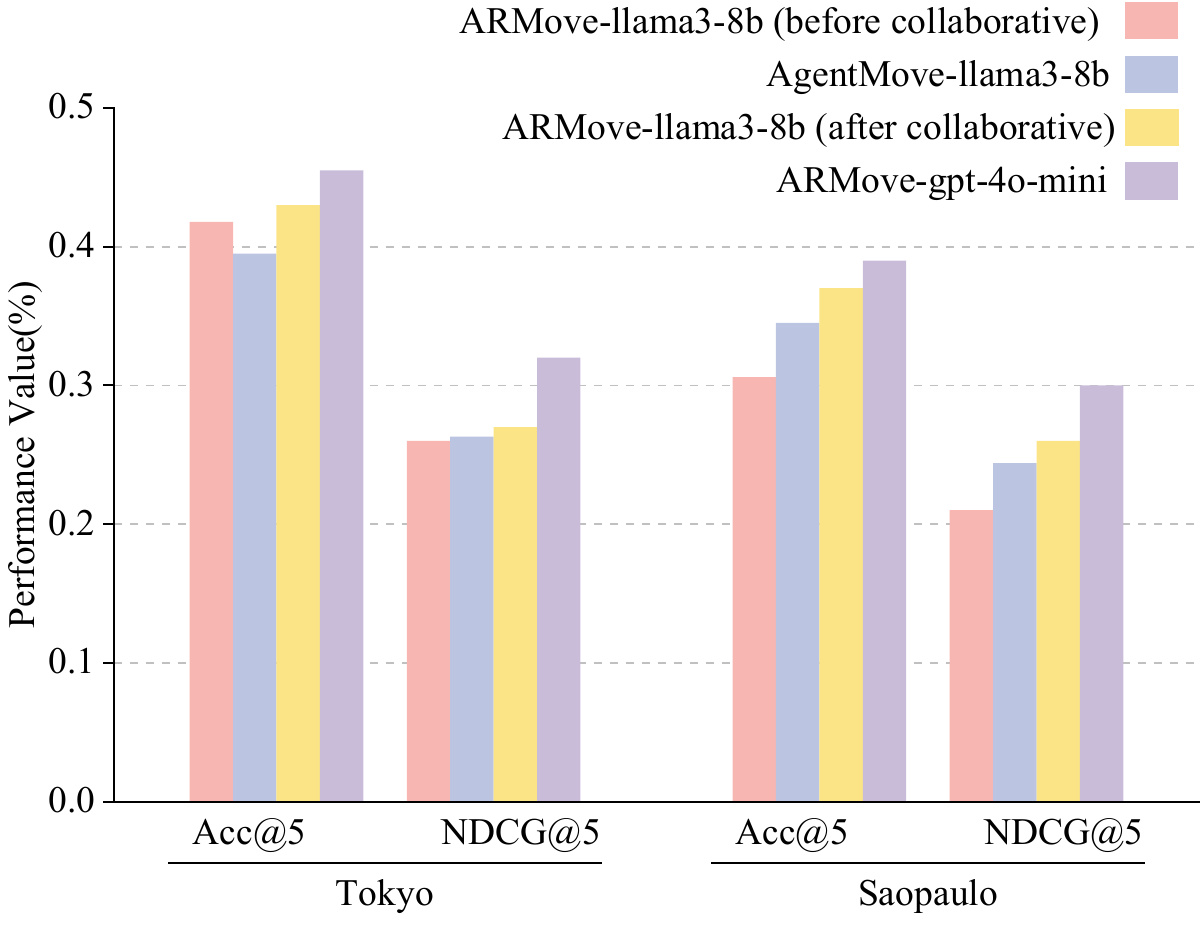}
    \caption{Fusion strategy of large and small models--using GPT-4o-mini in collaboration with Llama3-8B}
    \label{fig:3-2-l-s_pic}
\end{figure}

\begin{figure}[htbp]
  
    \centering
    \includegraphics[width=0.45\textwidth]{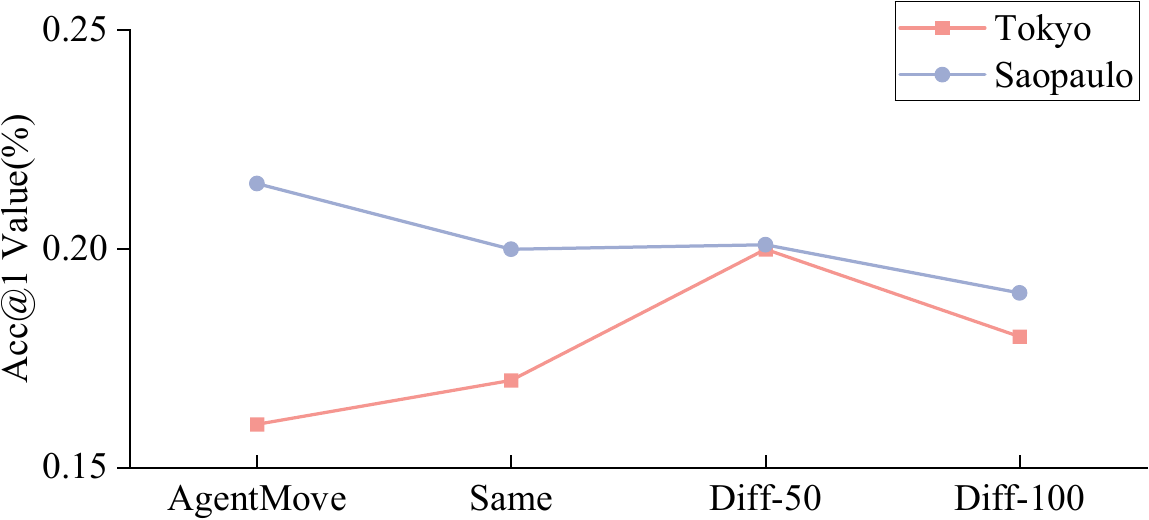}
    \caption{User Transfer Acc@5.}
    \label{fig:3-3-user_trans_pic}
\end{figure}

\textbf{Ablation Studies of Iteration Mechanism.} We compared model performance using 3, 5, and 10 iterations. The results consistently show that 5 iterations yields the best overall performance. Specifically, 5 iterations significantly boosted Acc@1 compared to 3 and 10 iterations: by 76\% and 60\% on the Moscow dataset, and by 32\% and 22\% on the Shanghai dataset. On the Shanghai dataset, Acc@5 also improved by 13\% and 14.3\%, respectively. These findings demonstrate that an iteration count of 5 is optimal across diverse datasets.

\textbf{Ablation Studies of User Grouping Strategy.} We compared three approaches: OL1 (initial user profile construction), L1L2 (further exploration of user interests), and L1L2M (fine-grained category merging). The results show that, after sequentially adding these modules, Acc@5 on all four datasets improved significantly, with gains ranging from 1\% to 19.25\%. This demonstrates the effectiveness of progressively refining user profiles and interest modeling.

\textbf{Ablation Studies of Feature Selection Module}. We evaluated three strategies: using only the large model for feature selection, combining the large model with all newly generated features, and a balanced weighting approach. Experimental results indicate that the balanced weighting method consistently outperforms the others across all 12 evaluation metrics in the four cities, compared to using only the large model for feature selection. More detailed experimental results can be found in the appendix.

\subsection{Transferability Evaluation}
The transfer experiments are mainly divided into three parts: collaborative transfer between large and small models, transfer between different users within the same city, and city fusion and direct transfer.

\textbf{Large-Small Model Collaboration.}As Figure~\ref{fig:3-2-l-s_pic} shows, we evaluate the effectiveness of knowledge transfer from our LLM-base (GPT-4o-mini) to a smaller model (Llama3-8B). While the guided Llama3-8B cannot fully surpass GPT-4o-mini—due to inherent capacity limitations—the guidance clearly yields improvements. Specifically, the guided ARMove outperforms the unguided version in 4 out of 6 metrics across two cities. Moreover, in 4 metrics, it even exceeds the performance of our baseline, AgentMove.

\textbf{User Transfer.}
As Figure~\ref{fig:3-3-user_trans_pic} shows, we evaluated the user-level transferability within the same city using the Tokyo and Moscow datasets as examples, comparing results after randomly replacing 50 and 100 users against the original set and the baseline. Overall, ARMove maintained stable performance, demonstrating strong generalization. For instance, on the Tokyo dataset, replacing 50 users increased Acc@1 from 0.17 to 0.20 (a 17.64\% improvement). Even after replacing 100 users, Acc@1 stabilized at 0.18 (a 5.88\% gain). Other metrics (Acc@5, NDCG@5) also remained stable or slightly improved, confirming ARMove's good robustness in intra-city user transfer scenarios.

\begin{figure}[htbp]
  \centering
  \includegraphics[width=0.45\textwidth]{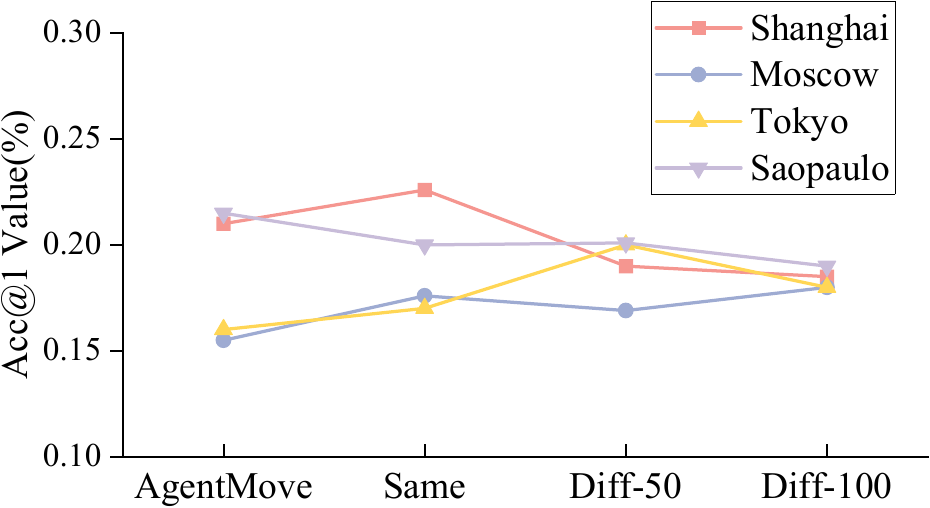}
  \caption{City transfer. `4C' denotes the integration of 200 users from four cities.}
  \label{fig:3-4-city-trans_pic}
\end{figure}

\begin{figure*}[htbp]
  \centering
  \includegraphics[width=\textwidth]{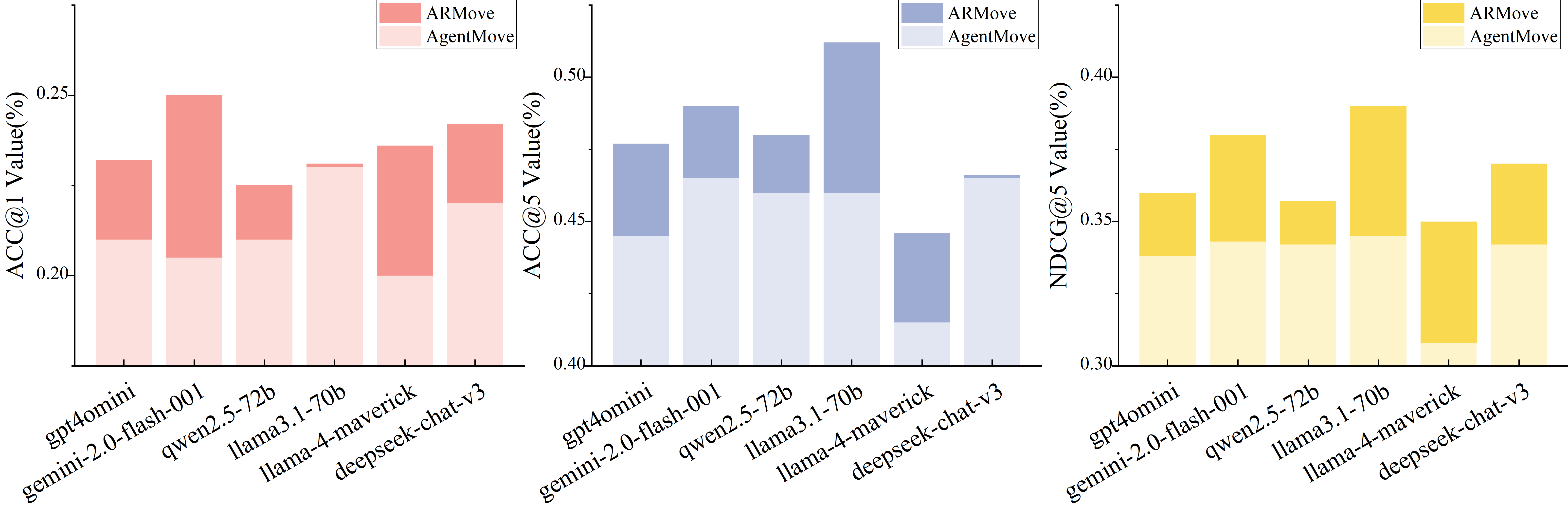}
  \caption{Incremental performance of ARMove on different models}
  \label{fig:3-5-multi-LLM-SH.png}
  \vspace{-0.5cm}
\end{figure*}

\textbf{City Fusion Transfer.} Multi-City Fusion Transfer: We trained the model using 200 randomly selected users fused from four cities and tested it on 200 users from a single target city. While performance was slightly lower than single-city training, this strategy is generally stable and offers a promising solution for data-scarce cities.
Direct City-to-City Transfer: This method showed particularly effective transferability between cities with similar characteristics. Notably, transferring knowledge from Moscow to Tokyo increased Acc@5 by $2.19\%$ compared to training solely on Tokyo data. Transfer performance on the Shanghai dataset was unsatisfactory, primarily due to its shorter collection period and distribution shifts. Excluding Shanghai, cross-city transfer remains stable, validating our multi-city fusion strategy. Detailed results are in the appendix and Figure~\ref{fig:3-4-city-trans_pic}.

\subsection{LLM Influence}
To further validate the applicability and generalization capability of our framework on various LLMs, we selected 6 recent models as well as models with larger parameter sizes for comparison. In Figure~\ref{fig:3-5-multi-LLM-SH.png}, we present the results in a stacked bar chart. It is evident that ARMove consistently outperforms the baseline AgentMove across all compared models, achieving higher performance on each respective model.

\section{Related Work} \label{sec:related}

\subsection{Human Mobility Prediction}
Research on mobility prediction task can be divided into two branch. The first branch is from the physics community where researchers try to find and define simple physical models that can capture the underlying mobility patterns behind the human behaviors~\cite{rendle2010factorizing, cheng2013you,du2025camscitygptpoweredagenticframework,feng2025urbanllavamultimodallargelanguage,wang2025mitigatinggeospatialknowledgehallucination,feng2025citygptempoweringurbanspatial,du2025trajagentllmagentframeworktrajectory}. However, with explainable characteristics, this branch research can only achieve limited performance in prediction accuracy. Another branch is from the computer science where researchers try to design model that can learn from the data directly. Due to the promising prediction accuracy, this kind of research becomes the mainstream in the past decades~\cite{liu2016predicting, feng2018deepmove, sun2020go, luo2021stan}. However, such research is highly dependent on training data and faces limitations~\cite{he2025mintqamultihopquestionanswering}, 
including a lack of transferability and interpretability. 

\subsection{Large Language Models and Agents}
Recently, large language models (LLMs)~\cite{achiam2023gpt, dubey2024llama} have achieved rapid development and making significant progress in many challenging fields like reasoning~\cite{wei2022cot,shang2024defint, roziere2023code} and mathematics~\cite{luo2023wizardmath}. However, due to the limitation of LLM in domain knowledge, it can be struggled in some real world tasks~\cite{Yang_2022,Long2024AUM,lei2025spider20evaluatinglanguage,zhang2025aflowautomatingagenticworkflow,zhang2025cybenchframeworkevaluatingcybersecurity,sridhar2025regentretrievalaugmentedgeneralistagent}. To address these challenges, LLM based agent framework~\cite{wang2024survey, shao2024beyond,kobalczyk2025activetaskdisambiguationllms,chen2024internetagentsweavingweb,chen2025spabenchcomprehensivebenchmarksmartphone,qian2025scalinglargelanguagemodelbased,yue2025synergisticmultiagentframeworktrajectory,zhou2025sweetrltrainingmultiturnllm,zhang2025agentmodelsinternalizingchainofaction,wu2025agenticreasoningstreamlinedframework} are proposed and succeed in unleashing the power of LLMs in challenging real-world tasks,  ChatDev~\cite{qian2023communicative} for software development and collective programming, AgentS~\cite{agashe2024agent} for automated computer operation, and LLMLight~\cite{lai2023llmlight} for traffic light control. Researchers also try to apply LLMs in the mobility prediction~\cite{wang2023would, beneduce2024large, feng2025agentmovelargelanguagemodel}. LLMMob~\cite{wang2023would} is the first work for applying LLM in mobility prediction. To solve the limitation in capturing mobilty patterns, AgentMove~\cite{feng2025agentmovelargelanguagemodel} build the first agentic framework for mobility prediction. 

\section{Conclusion}
In this paper, we introduce ARMove, a multi-agent framework powered by agentic reasoning for generalized human mobility prediction across global geolocations. ARMove incorporates a hierarchical user profiling agent, a novel feature mining agent, and a feature management agent. Furthermore, we propose a transfer module that leverages large models to guide small models, thereby effectively reducing computational costs. Extensive experiments on 12 metrics across four cities demonstrate ARMove's superiority, generalizability, and transferability in human mobility prediction tasks, while providing interpretable predictions through agentic reasoning. We plan to explore more efficient and cost-effective approaches by utilizing labeled data generated by large models to further guide the learning of small models, and to mine more multi-dimensional information of human mobility behaviors with large models. We also aim for the application of our framework to other spatiotemporal mining tasks. We believe that multi-agent systems based on agentic reasoning, such as ARMove, possess great potential in explaining why users move to predicted locations, bringing new perspectives to mobility behavior prediction.

 
\bibliography{aaai2026}

\appendix
\section{Appendix}\label{sec:appendix}

The tables here correspond to the ablation and transfer experiments discussed in the main text.

As shown in Table~\ref{table:app1}, Iteration+3, +5, and +10 represent setting the number of iterations to 3, 5, and 10, respectively, and updating weights according to different iteration counts. 
As illustrated in Table~\ref{table:app2}, ``only-LLM'' refers to user grouping performed solely by the large model, ``LLM+L1+L2'' adds first- and second-level groupings, and ``LLM+L1+L2+Merge'' further incorporates a small group merging strategy. 
As shown in Table~\ref{table:app3}, ``FS-only-llm'' indicates feature selection using only the large model, ``FS+llm+new feature'' adds new features provided by the large model, and ``FS+llm+new feature+weight-select'' adds weight balancing. 
As shown in Table~\ref{table:app4}, the main results demonstrate how the large model guides the small model.

\begin{table}[!ht]
\caption{Ablation Studies of Iteration}
\label{table:app1}
\centering
\resizebox{0.95\columnwidth}{!}{%
\begin{tabular}{lcccccc}
\Xhline{0.7pt}
\hline
\textbf{Models} & \multicolumn{3}{c}{\textbf{Shanghai\_isp}} & \multicolumn{3}{c}{\textbf{Tokyo}} \\
 & \textbf{acc@1} & \textbf{acc@5} & \textbf{NDCG@5} & \textbf{acc@1} & \textbf{acc@5} & \textbf{NDCG@5} \\ \hline
\textbf{Automove} & \textbf{0.226} & \textbf{0.477} & \textbf{0.360} & 0.170 & \textbf{0.455} & \textbf{0.320} \\
\textbf{Iteration(+3)} & 0.170 & 0.420 & 0.310 & \textbf{0.179} & 0.451 & 0.320 \\
\textbf{Iteration(+5)} & \textbf{0.226} & \textbf{0.477} & \textbf{0.360} & 0.170 & \textbf{0.455} & \textbf{0.320} \\
\textbf{Iteration(+10)} & 0.185 & 0.417 & 0.310 & 0.170 & 0.450 & 0.300 \\ \hline
\textbf{Models} & \multicolumn{3}{c}{\textbf{Moscow}} & \multicolumn{3}{c}{\textbf{Saopaulo}} \\
 & \textbf{acc@1} & \textbf{acc@5} & \textbf{NDCG@5} & \textbf{acc@1} & \textbf{acc@5} & \textbf{NDCG@5} \\ \hline
\textbf{Automove} & \textbf{0.176} & 0.383 & \textbf{0.293} & \textbf{0.200} & \textbf{0.390} & \textbf{0.300} \\
\textbf{Iteration(+3)} & 0.100 & 0.375 & 0.240 & 0.200 & 0.370 & 0.290 \\
\textbf{Iteration(+5)} & \textbf{0.176} & 0.383 & \textbf{0.293} & \textbf{0.200} & \textbf{0.390} & \textbf{0.300} \\
\textbf{Iteration(+10)} & 0.110 & \textbf{0.390} & 0.250 & 0.200 & 0.390 & 0.300 \\ \hline \Xhline{0.7pt}
\end{tabular}%
}
\end{table}

\begin{table}[!ht]
\caption{Ablation Studies of User Grouping}
\label{table:app2}
\centering
\resizebox{0.95\columnwidth}{!}{%
\begin{tabular}{lcccccc}
\Xhline{0.7pt}
\hline
\textbf{Models} & \multicolumn{3}{c}{\textbf{Shanghai\_isp}} & \multicolumn{3}{c}{\textbf{Tokyo}} \\
 & \textbf{acc@1} & \textbf{acc@5} & \textbf{NDCG@5} & \textbf{acc@1} & \textbf{acc@5} & \textbf{NDCG@5} \\ \hline
\textbf{Automove} & \textbf{0.226} & \textbf{0.477} & 0.360 & 0.170 & \textbf{0.455} & 0.320 \\
\textbf{Only-LLM} & 0.170 & 0.400 & 0.290 & 0.210 & 0.440 & \textbf{0.340} \\
\textbf{LLM+L1+L2} & 0.160 & \textbf{0.400} & 0.290 & \textbf{0.200} & 0.450 & 0.330 \\
\textbf{LLM+L1+L2+Merge} & \textbf{0.226} & \textbf{0.477} & \textbf{0.360} & 0.170 & \textbf{0.455} & 0.320 \\ \hline
\textbf{Models} & \multicolumn{3}{c}{\textbf{Moscow}} & \multicolumn{3}{c}{\textbf{Saopaulo}} \\
 & \textbf{acc@1} & \textbf{acc@5} & \textbf{NDCG@5} & \textbf{acc@1} & \textbf{acc@5} & \textbf{NDCG@5} \\ \hline
\textbf{Automove} & \textbf{0.176} & \textbf{0.383} & \textbf{0.293} & \textbf{0.200} & \textbf{0.390} & \textbf{0.300} \\
\textbf{Only-LLM} & 0.130 & 0.380 & 0.270 & 0.190 & 0.370 & 0.290 \\
\textbf{LLM+L1+L2} & 0.090 & 0.370 & 0.240 & 0.190 & 0.370 & 0.290 \\
\textbf{LLM+L1+L2+Merge} & 0.176 & \textbf{0.383} & \textbf{0.293} & \textbf{0.200} & \textbf{0.390} & \textbf{0.300} \\
\hline \Xhline{0.7pt}
\end{tabular}%
}
\end{table}

\begin{table}[!ht]
\caption{Ablation Studies of Feature Selection}
\label{table:app3}
\centering
\resizebox{0.95\columnwidth}{!}{%
\begin{tabular}{lllllll}
\hline
\Xhline{0.8pt}
\multirow{2}{*}{\textbf{Models}} & \multicolumn{3}{c}{\textbf{Shanghai\_isp}} & \multicolumn{3}{c}{\textbf{Tokyo}} \\
 & \textbf{acc@1} & \textbf{acc@5} & \textbf{NDCG@5} & \textbf{acc@1} & \textbf{acc@5} & \textbf{NDCG@5} \\ \hline
\textbf{Automove} & \ul{\textbf{0.226}} & 0.477 & \textbf{0.36} & 0.170 & \textbf{0.455} & 0.32 \\
\textbf{FS-only-llm} & 0.226 & 0.472 & 0.36 & \textbf{0.184} & 0.446 & \textbf{0.32} \\
\textbf{FS+llm+new feature} & 0.175 & \textbf{0.41} & 0.31 & \textbf{0.169} & 0.43 & 0.30 \\
\textbf{FS+llm+new feature+weight\_select} & \ul{0.226} & 0.477 & \textbf{0.360} & 0.170 & \textbf{0.455} & 0.320 \\ \hline
\multirow{2}{*}{\textbf{Models}} & \multicolumn{3}{c}{\textbf{Moscow}} & \multicolumn{3}{c}{\textbf{Saopaulo}} \\
 & \textbf{acc@1} & \textbf{acc@5} & \textbf{NDCG@5} & \textbf{acc@1} & \textbf{acc@5} & \textbf{NDCG@5} \\ \hline
\textbf{Automove} & 0.176 & 0.383 & \textbf{0.293} & \textbf{0.200} & \textbf{0.390} & \textbf{0.300} \\
\textbf{FS-only-llm} & \textbf{0.178} & 0.382 & 0.280 & 0.175 & 0.360 & 0.270 \\
\textbf{FS+llm+new feature} & 0.100 & \textbf{0.385} & 0.250 & 0.190 & 0.378 & 0.290 \\
\textbf{FS+llm+new feature+weight\_select} & 0.176 & 0.383 & \textbf{0.293} & \textbf{0.200} & \textbf{0.390} & \textbf{0.300} \\ \hline \Xhline{0.7pt}
\end{tabular}%
}
\end{table}

\begin{table}[!ht]
\caption{Large-small model collaboration strategy}
\label{table:app4}
\centering
\resizebox{0.95\columnwidth}{!}{%
\begin{tabular}{lllllll}
\hline
\Xhline{0.7pt}
\textbf{Large and Small models} & \multicolumn{3}{c}{\textbf{Shanghai\_isp}} & \multicolumn{3}{c}{\textbf{Moscow}} \\
 & \textbf{acc@1} & \textbf{acc@5} & \textbf{NDCG@5} & \textbf{acc@1} & \textbf{acc@5} & \textbf{NDCG@5} \\ \hline
\textbf{automove-gpt-4o-mini} & \textbf{0.226} & \textbf{0.477} & \textbf{0.36} & \textbf{0.176} & \textbf{0.383} & \textbf{0.293} \\
{\color[HTML]{080F17}\textbf{agentmove-llama3-8b}} & {\color[HTML]{080F17}\ul{0.175}} & {\color[HTML]{080F17}\ul{0.45}} & {\color[HTML]{080F17}\ul{0.3171}} & {\color[HTML]{080F17}0.115} & {\color[HTML]{080F17}\ul{0.32}} & {\color[HTML]{080F17}\ul{0.2218}} \\
\textbf{ARmove-llama3-8b-no-transfer} & 0.10 & 0.44 & 0.25 & \ul{0.12} & 0.30 & 0.21 \\
\textbf{ARmove-llama3-8b-tranfer} & 0.11 & 0.437 & 0.29 & {\color[HTML]{080F17}\ul{0.12}} & 0.30 & 0.21 \\ \hline
 & \multicolumn{3}{c}{\textbf{Tokyo}} & \multicolumn{3}{c}{\textbf{Saopaulo}} \\
 & \textbf{acc@1} & \textbf{acc@5} & \textbf{NDCG@5} & \textbf{acc@1} & \textbf{acc@5} & \textbf{NDCG@5} \\ \hline
\textbf{automove-gpt-4o-mini} & \textbf{0.170} & \textbf{0.455} & \textbf{0.32} & \textbf{0.200} & \textbf{0.390} & \textbf{0.300} \\
\textbf{agentmove-llama3-8b} & \ul{0.115} & 0.395 & 0.2632 & 0.135 & 0.345 & 0.2440 \\
\textbf{ARmove-llama3-8b-no-transfer} & 0.096 & 0.418 & 0.26 & \ul{0.19} & 0.306 & 0.21 \\
\textbf{ARmove-llama3-8b-Transfher} & 0.090 & \ul{0.430} & \ul{0.27} & 0.13 & \ul{0.37} & \ul{0.26} \\ \hline \Xhline{0.7pt}
\end{tabular}%
}
\end{table}

\end{document}